# THE SNS CRYOGENIC CONTROL SYSTEM: EXPERIENCES IN COLLABORATION [1]

W. H. Strong, P. A. Gurd, SNS ORNL, Oak Ridge, TN 37830 USA
J. D. Creel, B. S. Bevins, TJNAF, Newport News, VA 23606 USA

Abstract

The cryogenic system for the Spallation Neutron Source (SNS) is designed by Jefferson Laboratory (JLab) personnel and is based on the existing JLab facility. Our task is to use the JLab control system design [2] as much as practical while remaining consistent with SNS control system standards. Some aspects of the systems are very similar, including equipment to be controlled, the need for PID loops and automatic sequences, and the use of EPICS. There are differences in device naming, system hardware, and software tools. The cryogenic system is the first SNS system to be developed using SNS standards. This paper reports on our experiences in integrating the new and the old.

## 1 PROJECT DESCRIPTION

The SNS project is a partnership involving six DOE national laboratories: Argonne, Brookhaven, Jefferson, Lawrence Berkeley, Los Alamos, and Oak Ridge. The project is described in more detail on the SNS website and elsewhere in these proceedings. [3, 4]

The cryogenic system provides the refrigeration necessary to supply the 2.1° Kelvin liquid helium required for operation of the SNS accelerator superconducting cavities. Once the cryogenic refrigerator system is placed in service, it is expected to run continuously. Shutting down the cryogenic system is very expensive due to the loss of beam time and helium. Cryogenic refrigerator systems operate best with a fairly constant load. There are very few independent control loops; a change to any one control variable affects the entire system. The Cryogenic Control System (CCS) provides the coordinated control algorithms to support the complex operation of the cryogenic refrigeration system.

## 2 CONTROL SYSTEM OVERVIEW

The top-level Integrated Control System (ICS) for the entire SNS project is based on the Experimental Physics and Industrial Control System (EPICS) suite of monitoring and control software. For the cryogenic systems, EPICS provides the operator interface for control and monitoring, alarming and archiving of selected signals, and routines to analyze real-time and historic data. All high level control algorithms, control loops and automatic sequences reside in EPICS. EPICS routines constantly monitor the state of the cryogenic system and update control outputs to keep the refrigerator system stable and the load constant. Programmable Logic Controllers (PLC) are used to interface signals from sensors and actuators to the EPICS Input/Output Controllers (IOC). PLCs perform low level interlocks, control loops and sequences.

## 3 DESIGN COLLABORATION

JLab is designing and procuring the cryogenic plant and cryomodules including all mechanical equipment, instrumentation, and control devices. The control system at JLab uses EPICS. The SNS CCS duplicates the JLab system as much as practicable.

The SNS Integrated Control System Group at the Oak Ridge National Laboratory (ORNL) is producing the CCS from the sensors and actuators up. This includes cables, signal conditioning, PLCs, VME based IOCs, boot servers, operator interface workstations, network hardware, and all control room equipment.

Los Alamos National Laboratory (LANL) provides general EPICS support. LANL also provides the RF control system for the cryomodules and the RF signals needed by the CCS for use in heater control.

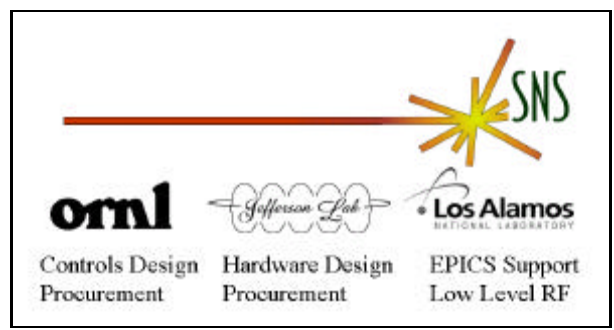

Figure 1: Collaboration on the SNS Cryogenic System.

Due to advances in hardware and software, there are differences in the CCS implementation between JLab and SNS. JLab has used their experience with an operating cryogenic system to provide invaluable advice and guidance in the selection of hardware and software components for use by SNS.

## 4 HARDWARE

### 4.1 Cryogenic Control Room Hardware

The SNS Control Room will be used for control of all accelerator systems. A small local control room will be included in the Central Helium Liquefier building for operation of the cryogenic system. It will house the PC-based computer workstations, network hardware, and dedicated instrumentation for efficient control of the cryogenic refrigerator system. Two workstations will be used for operator interface functions and IOC boot servers. These two machines will run Linux and contain all source code and libraries necessary to operate and maintain the CCS in the event that the communication link to the SNS integrated control system server is lost. They will be configured such that if the primary boot server fails, the backup can assume the role of boot server with minimal operator intervention. These machines will provide the resources required for archiving, alarming and coordination of system control among the 10 cryogenic system IOCs. Each workstation will have 3 monitors for operator use.

A third Linux workstation with a single monitor will be provided for engineering and software development activities. These workstations will be identical so that they are interchangeable.

A Windows based workstation with a single monitor will be provided for PLC software development and maintenance.

While the hardware is different, this control room configuration is functionally very similar to the cryogenic control room at JLab. The design is based on the existing JLab cryogenic control room with updates to incorporate improvements suggested by JLab control system experts.

### 4.2 Cryogenic Control System Network

The SNS ICS network is divided into several subnets. The CCS subnet is separated from and independent of the remainder of the network. It includes redundant central switches and interconnecting fiber optic cables. [5] Experience at JLab determined that separation of the CCS network from the remainder of the network was necessary to provide the high availability required of the CCS.

Some heat load parameters needed for optimum operation of the refrigerator system originate outside the CCS subnet in the Low Level RF (LLRF) subsystem. If communication with LLRF is lost, the CCS inhibits beam operation and automatically takes the facility to a safe operational state.

### 4.3 Signal Interface Hardware

The major differences between the JLab and SNS CCS are in the interface hardware. VME/VXI crates with a MVME2101 power PC are used for the SNS IOCs. Custom utility modules provide crate information such as power supply status, crate temperature, etc. Custom timing modules are used to time synchronize all IOCs to a master clock so that all data can be time correlated. Some VME input modules are used for special sensors:
- Highland Technology V460 for silicon diode cryogenic temperature sensors
- Highland Technology V550 for linear variable differential transformer (LVDT) position sensors

Where possible, SNS uses the VME version of JLab CAMAC modules. Since no VME version of the helium liquid level module is available, custom signal conditioning is required at SNS.

Allen-Bradley ControlLogix PLC standard input and output modules are used to interface most sensors and actuators that require analog and discrete signals.

### 4.4 Local Control Hardware

Local control must be provided for some process valves so that the cryogenic system can continue to operate in the event of loss of an IOC. For the JLab CCS, local control is provided by custom valve control modules in the CAMAC crates. The SNS ICS provides Allen-Bradley PanelView terminals directly connected to the PLCs to fulfill this requirement. The PanelView at SNS provides additional capability by displaying many of the process parameters that would be used in the valve control loops in the IOC.

## 5 SOFTWARE

SNS global controls have mandated a number of software standards to be used in all components. These include the use of EPICS, device naming conventions, the use of the Oracle database and operator screen color conventions.

### 5.2 EPICS

The use of EPICS makes available a large body of software and support. Since JLab also uses EPICS, some JLab software such as the LVDT device driver (used for position measurements) can be used directly. Other software can be substantially copied; the V460 device driver (used to measure cryogenic temperatures) was adapted from the LVDT driver. The CPID record, used extensively in the CCS, is a part of EPICS base created by JLab for their control system. EPICS facilitates the SNS collaboration; SNS collaborators at

Los Alamos National Laboratory developed the Ethernet Industrial Protocol driver that provides the communication link between the IOC and the PLC.

### 5.3 Device Naming and Databases

A common naming scheme is crucial to integrate signals among the systems and collaborating laboratories. SNS cryogenic controls personnel have used a Microsoft Access application as a step toward the development of a project-wide database. The lessons learned in the process of creating EPICS databases from this application are being incorporated into the project-wide ORACLE technical database. Examples of all cryogenic system device types have been included in the test database. So far, the records created have very few links to other records. Graphical tools will be required to create and maintain highly linked databases, for example those involving calculations or PID loops. Because many of the cryogenic signals and calculations are adapted from JLab, test versions have been created using the same tool that JLab uses, CapFast. A general SNS solution to this problem is still required. Alarms and archive configuration files will be produced from the project-wide database. In the interim, these must be produced from our local database.

### 5.4 Display Manager

SNS has selected the Extensible Display Manager (EDM), which is maintained at ORNL. A script was created to translate the JLab operator screens, which run under MEDM, into EDM. After the screens were translated, the SNS color and font standards were applied. The screens were updated to reflect the SNS plant hardware design and incorporate improvements suggested by JLab. Finally, the JLab device names were replaced by the SNS standard device names. Although the final versions are substantially different from the original JLab screens, this process provided a good starting point for much of the operator interface. This process has worked for the main compressor screen and first and second stage compressor screens. Some features have been added to EDM for SNS: symbolic colors, symbolic font declarations, undo and a striptool widget, as well as widgets to display parts of SNS-standard names.

### 5.5 Control Algorithms

Both SNS and JLab use EPICS or routines accessed through EPICS for all high-level control loops and algorithms. An example is the cavity heater control algorithm.

A cryogenic RF cavity operates at a constant pressure. Cryogenic refrigerator systems operate best with a constant load. The flow, temperature, and pressure in the common primary return line from the cryogenic cavities are factors in the load. The return pressure, which must be held constant for proper RF cavity operation, is closely related to the heat absorbed by the helium liquid in each of the 81 superconducting cavities and the return flow. One of the major dynamic heat sources is RF loss into the cavity. When RF to a cavity is turned off, power to a resistance heater in the cavity is increased to keep the heat load constant. All cavities share a common return line, and a change in one cavity affects all other cavities to some extent. If too much total heat is applied, the return pressure will increase and the return flow will try to increase. The return flow control loop will react to keep the flow constant. Coordination of all heaters is required to keep the total heat load and return flow balanced. An algorithm running on the boot server in the cryogenic control room provides the coordination of heater control across all cavities. JLab, LANL, and SNS personnel are working together to develop this control algorithm and provide all required input parameters.

## 6 SUMMARY

Collaboration among several partners over great distances is working. The SNS project benefits by having access to the experience of the personnel at the partner laboratories.

## REFERENCES

[1] Work supported by the US Department of Energy under contract DE-AC05-00OR22725
[2] Marie S. Keesee and Brian S. Bevins, "The CEBAF Control System for the CHL", /Advances in Cryogenic Engineering/, Vol. 41, pp. 649-654, Plenum Press, New York, 1996
[3] SNS Website  http://www.sns.gov/
[4] Dave Gurd, TUDT002 "Management of a Large, Distributed Control System Development Project", ICALEPCS 2001
[5] Bill DeVan, TUAP055 "Plans for the Spallation Neutron Source Integrated Control System Network", ICALEPCS 2001